%% file: main.tex
\documentclass[sigconf]{acmart}
\setcopyright{none}
\renewcommand\footnotetextcopyrightpermission[1]{} 
\AtBeginDocument{%
  \providecommand\BibTeX{{%
    \normalfont B\kern-0.5em{\scshape i\kern-0.25emb}\kern-0.8em\TeX}}}

\usepackage{microtype}
\usepackage{times}
\usepackage{latexsym}
\usepackage{url}
\usepackage{multirow}
\usepackage{balance}
\usepackage{latexsym}
\usepackage{amsmath}
\usepackage{paralist}
\usepackage{makecell}
\usepackage{hyperref}
\usepackage{subcaption}
\usepackage{caption}
\usepackage{graphicx}
\usepackage{url}
\usepackage{multicol}
\usepackage{multirow}
\usepackage{booktabs}
\usepackage{color}
\usepackage{transparent}
\usepackage{array}
\usepackage{balance}
\usepackage{bm}
\definecolor{midnightgreen}{rgb}{0.0, 0.29, 0.33}
\definecolor{darkpink}{rgb}{0.91, 0.33, 0.5}

\makeatletter
\renewcommand\@formatdoi[1]{\ignorespaces}
\makeatother

\newcommand{\myparagraph}[1]{\paragraph*{\hspace*{-\parindent}\normalsize{\bf{#1}}}}

\begin{document}

\title{CMT in TREC-COVID Round 2: Mitigating the Generalization Gaps from Web to Special Domain Search}

\author{Chenyan Xiong$^{^\spadesuit*}$, Zhenghao Liu$^{\heartsuit*}$, Si Sun$^{\heartsuit*}$, Zhuyun Dai$^{\clubsuit*}$, Kaitao Zhang$^{\heartsuit*}$, Shi Yu$^{\heartsuit}$}\authornote{\ \ indicates equal contribution.}
\author{Zhiyuan Liu$^\heartsuit$,  Hoifung Poon$^\spadesuit$, Jianfeng Gao$^\spadesuit$, Paul Bennett$^\spadesuit$}
\affiliation{Tsinghua University$^\heartsuit$, Microsoft Research$^\spadesuit$, Carnegie Mellon University$^\clubsuit$
} 
\affiliation{
\texttt{\{liu-zh16, s-sun17, zkt18, yus17\}@mails.tsinghua.edu.cn};\\
\texttt{liuzy@tsinghua.edu.cn}; \texttt{zhuyund@cs.cmu.edu}; \\
\texttt{\{chenyan.xiong, hoifung, jfgao, pauben\}@microsoft.com}
}

\input{Sections/0_Abstract}

\pagestyle{plain}
\maketitle

\input{Sections/1_Introduction}
\input{Sections/2_DataStudy}
\input{Sections/3_Methodology}
\input{Sections/4_Experiment}
\input{Sections/5_Evaluation}

\input{Sections/6_NegativeRes}

\myparagraph{Acknowledgments} We thank Luyu Gao for sharing the implementation of Dense Retrieval, the Track organizers for hosting this track, Sean Macavaney for releasing the medical MS MARCO filter, and Jimmy Lin \& the Anserini project for open sourcing the well-rounded BM25 first stage retrieval.
\balance

\bibliographystyle{ACM-Reference-Format}
\bibliography{sample-base}

\end{document}

%% file: Sections/0_Abstract.tex
\begin{abstract}
Neural rankers based on deep pretrained language models (LMs) have been shown to improve many information retrieval benchmarks. However, these methods are affected by their the correlation between pretraining domain and target domain and rely on massive fine-tuning relevance labels. Directly applying pretraining methods to specific domains may result in suboptimal search quality because specific domains may have domain adaption problems, such as the COVID domain. This paper presents a search system to alleviate the special domain adaption problem. The system utilizes the domain-adaptive pretraining and few-shot learning technologies to help neural rankers mitigate the domain discrepancy and label scarcity problems. Besides, we also integrate dense retrieval to alleviate traditional sparse retrieval's vocabulary mismatch obstacle. Our system performs the best among the non-manual runs in Round 2 of the TREC-COVID task, which aims to retrieve useful information from scientific literature related to COVID-19. Our code is publicly available at \url{https://github.com/thunlp/OpenMatch}.
\end{abstract}


%% file: Sections/1_Introduction.tex
\section{Introduction}
Recent years have witnessed continuous successes of neural ranking models in information retrieval~\cite{pang2017deeprank, dai2018convolutional, macavaney2019cedr, xiong2020approximate}. Most notably, deep pretrained language models (LMs) achieve state-of-the-art performance on several web search benchmarks~\cite{yang2019simple, nogueira2019passage, craswell2020overview}. Their success relies on the learned semantic information from general domain corpus with the language model pretraining~\cite{craswell2020overview, zhang2019generic}.

However, ranking models in specific domains usually face the domain adaption problem, which comes from two generalization gaps between the general and the specific domain. The first gap derives from the discrepancy of vocabulary distributions in different domains. Taking the COVID domain as an example~\cite{wang2020cord, voorhees2020trec}, the earliest related publication appeared at the end of 2019. Even pretrained LMs targeting the biomedical domain~\cite{Beltagy2019SciBERT, lee2020biobert} are unfamiliar with new medical terms like COVID-19 because their pretraining corpora have not contained such new terminologies. The other gap is the label scarcity. For the specific searching scenario, large-scale relevance labels are luxury, such as biomedical and scientific domains.

In addition, most information retrieval (IR) systems usually use sparse ranking methods in the first-stage retrieval, such as BM25, which are based on term-matching signals to calculate the relevance between query and document. Nevertheless, these systems may fail when queries and documents use different terms to describe the same meaning, which is known as the vocabulary mismatch problem~\cite{furnas1987vocabulary, croft2010search}. The vocabulary mismatch problem of sparse retrieval has become an obstacle to existing IR systems, especially for specific domains that have lots of in-domain terminologies. 


This paper presents a solution to alleviate the specific domain adaption problem with three core technics. The first one conducts domain-adaptive pretraining (DAPT)~\cite{gururangan2020don} to help pretrained language models learn semantics of special domain terminologies to keep the language knowledge is the latest. The second one uses Contrast Query Generation (ContrastQG) and ReInfoSelect~\cite{zhang2020selective} to mitigate the label scarcity problem in the specific domain. ContrastQG and ReInfoSelect focus on generating and filtering pseudo relevance labels to further improve ranking performance, respectively. Finally, our system integrates dense retrieval to alleviate the sparse retrieval's vocabulary mismatch bottleneck.
Dense retrieval can encode query and document to dense vectors to measure the relevance between query and document in the latent semantic space~\cite{karpukhin2020dense, gao2020complementing, luan2020sparse, chang2020pre,xiong2020approximate}.

Using above technologies, our system achieves the best performance among non-manual groups in Round 2 of TREC-COVID~\cite{voorhees2020trec}, which is a COVID-domain TREC task to evaluate information retrieval systems for searching COVID-19 related literature.

The next section will analyze the generalization gaps and vocabulary mismatch faced by the COVID domain search. Sec.\ref{sec:system} and Sec.\ref{sec:implement} describe in detail how our system alleviates these problems. Sec.\ref{sec:evaluation} shows the evaluation results and hyperparameter study. In the Sec.\ref{sec:attempts} and Sec.\ref{sec:concern}, we discuss the negative attempts and our concerns of the residual collection evaluation~\cite{salton1990improving} used in TREC-COVID.

%% file: Sections/2_DataStudy.tex
\section{Data Study}
This section studies the generalization gaps from web to COVID domain, and the vocabulary mismatch problem of sparse retrieval.

\textbf{Domain Discrepancy.} Most existing pretrained language models divide uncommon words into subwords, which aims to alleviate the out-of-vocabulary problem~\cite{sennrich2015neural}. As shown in Figure~\ref{fig:corpus_mismatch}, the subword ratio of TREC-COVID queries is dramatically higher than that of the web domain dataset, MS MARCO~\cite{bajaj2016ms}. The results show that existing pretrained language models treat most COVID-domain terminologies as unfamiliar words, indicating a considerable discrepancy between the existing pretraining and the COVID domain.

\textbf{Label Scarcity.} The label scarcity in the COVID domain search is very prominent. Only 30 queries were judged in the second round of TREC-COVID. In contrast, medical MS MARCO contains more than 78,800 annotated queries, which is the medical subset of MS MARCO filtered by the previous work~\cite{macavaney2020sledge}.

\textbf{Vocabulary Mismatch.} We observed that BM25 only covered 35\% of relevant documents in the top 100 retrieved documents. The result reveals that retrieving relevant documents only according to term-matching signals will hinder the search system's effectiveness.

\input{Diagrams/corpus_mismatch}

%% file: Diagrams/corpus_mismatch.tex
\begin{figure}[t]
\includegraphics[width=0.45\textwidth]{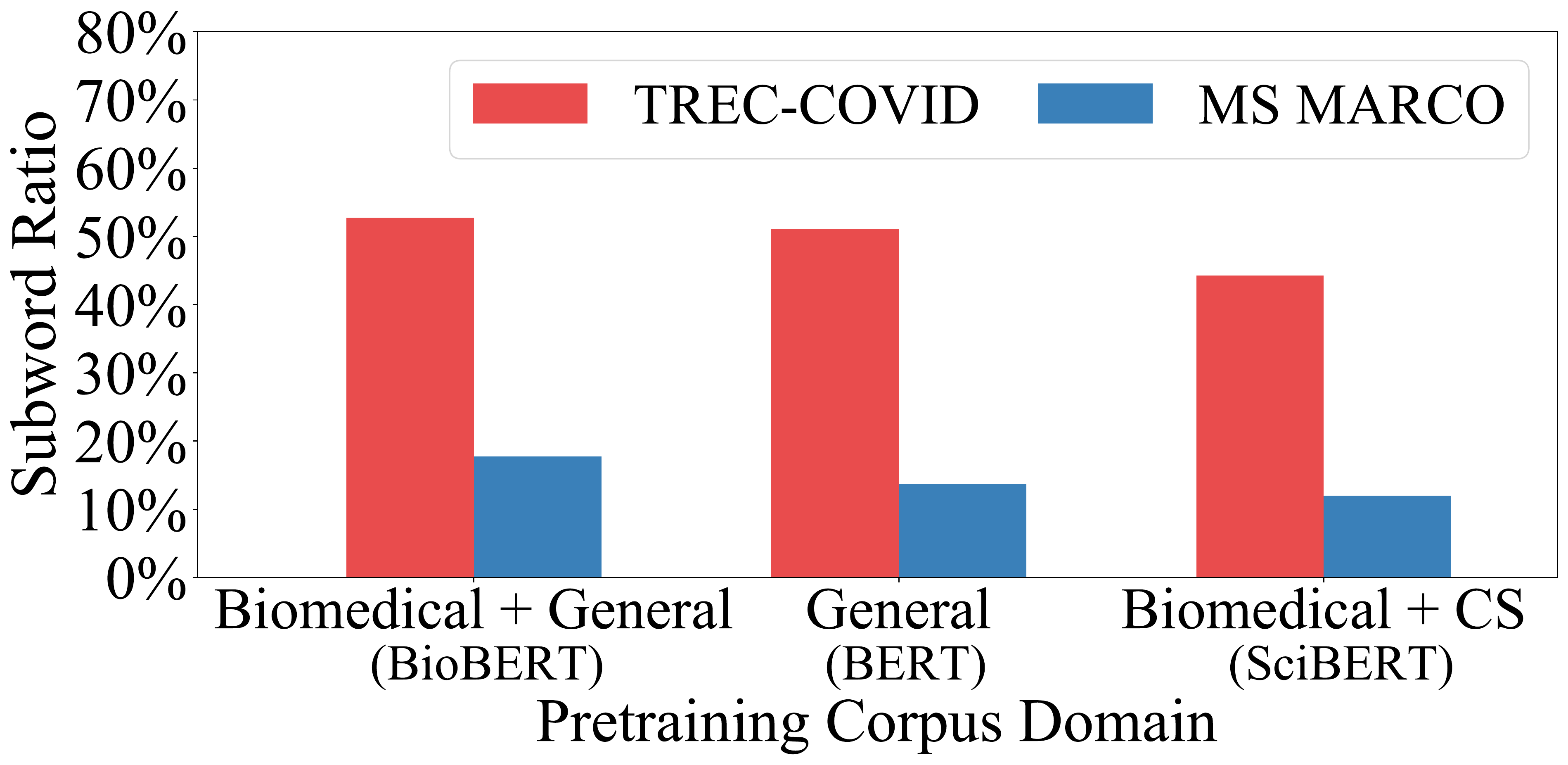}
\caption{\label{fig:corpus_mismatch}The proportion of query words that are decomposed into subwords by the pretrained language model's vocabulary.}
\end{figure}

%% file: Sections/3_Methodology.tex
\section{System Description}\label{sec:system}

Our system employs a two-stage retrieval architecture, which utilizes BM25 for base retrieval and SciBERT~\cite{Beltagy2019SciBERT} for reranking. The domain-adaptive pretraining and two few-shot learning techniques are used to mitigate the generalization gaps faced by SciBERT in the COVID domain. Dense retrieval is also incorporated into our system to alleviate BM25's vocabulary mismatch problem.

\subsection{Domain-Adaptive Pretraining}
SciBERT has been used in our system since it is pretrained with scientific texts and biomedical publications. However, COVID is a new concept that has not appeared in previous pretraining corpora. Therefore, we conduct domain-adaptive pretraining (DAPT)~\cite{gururangan2020don} for SciBERT. Our approach is straightforward to continuously train SciBERT with CORD-19 corpus~\cite{wang2020cord}, which is a growing collection of scientific papers about COVID-19 and coronavirus.

\subsection{Few-Shot Learning}
We introduce two few/zero-shot learning methods named \text{ContrastQG} and ReInfoSelect~\cite{zhang2020selective} to alleviate the label scarcity challenge when fine-tuning the neural ranking model. Specifically, we first use ContrastQG to generate weakly supervised data in a zero-shot manner and then utilize a weak supervision data selection method, ReInfoSelect, to recognize high quality training data.


\textbf{ContrastQG} is a zero-shot data synthetic method aiming to generate queries for synthesizing weakly supervised relevance signals. Unlike the prior work~\cite{ma2020zero}, ContrastQG synthesizes a query given a relevant text pair rather than a single related text, which can capture the specificity between two documents to generate more meaningful queries instead of keyword-style queries.


The entire synthesis process uses two query generators named $QG$ and $ContrastQG$, which aim to generate pseudo queries according to documents. Both $QG$ and $ContrastQG$ are implemented with standard GPT-2~\cite{radford2019language}. $QG$ is trained on medical MS MARCO's positive passage-query pairs ($d_+$, $q$) following the previous method~\cite{ma2020zero}. $ContrastQG$ is directly trained on medical MS MARCO's triples by encoding the concatenated text of positive and negative passages ($d_+$, $d_-$) to generate query $q$. 

At inference time, we first leverage $QG$ to generate queries $q$ based on a single COVID domain document $d$:
$$
q = QG(d).
$$
Then we utilize BM25 to retrieve two related documents ($d'_+$, $d'_-$) that show different correlation according to the generated query $q$. Finally, $ContrastQG$ is used to generate another query $q'$ based on the two contrastive documents ($d'_+$, $d'_-$):
$$
q' = ContrastQG(d'_+, d'_-).
$$
The synthetic triple $(q', d'_+, d'_-)$ is used as weakly supervised data to train the neural ranker.

\textbf{ReInfoSelect}~\cite{zhang2020selective} uses reinforcement learning to select weak supervision data. ReInfoSelect evaluates the neural ranker's performance on the target data and regards the NDCG difference as the reward. Then the reward signal from target data is propagated to guide data selector via the policy gradient. 

In our system, we use ContrastQG and medical MARCO to construct the weakly supervised data. The annotated data of TREC-COVID Round 1 is used as the target data. The trial-and-error learning mechanism of ReInfoSelect can select proper weakly supervised data according to neural ranker's performance in the target domain, which helps to further mitigate the domain discrepancy.

\input{Diagrams/overall}

\subsection{Dense Retrieval}
Dense retrieval maps queries and documents to the same distributed representation space and retrieves related documents based on the similarities between document vectors and query vectors~\cite{karpukhin2020dense,xiong2020approximate}. 

Let each training instance contain a query $q$, relevant (positive) document $d_{+}$ and $m$ irrelevant (negative) documents $D_{-} = \{d_{-}^{j}\}_{j=1}^{m}$. Dense retrieval first encodes the query $q$ and all documents $d$ to dense vectors $\boldsymbol{q}$ and $\boldsymbol{d}$. Then the similarity of $\boldsymbol{q}$ and $\boldsymbol{d}$ is calculated as $sim(\boldsymbol{q}, \boldsymbol{d})$. The training objective can be formulated as learning a distributed representation space that the positive document has a higher similarity to the query than all negative documents:
$$
loss(q, d_{+}, D_{-}) = -\text{log}\frac{e^{sim(\boldsymbol{q}, \boldsymbol{d_{+}})}}{e^{sim(\boldsymbol{q}, \boldsymbol{d_{+}})} + \sum_{j=1}^{m}e^{sim(\boldsymbol{q}, \boldsymbol{d_{-}^{j}})}},
$$
where the similarity $sim(\cdot, \cdot)$ is the dot product between vectors.



%% file: Diagrams/overall.tex
\begin{table*}[t]
    \centering
    \caption{Overall accuracy in Round 2 of TREC-COVID. The testing results of baselines and our three submitted runs (marked with asterisk$\text{}^*$) are from official evaluations. Compared baselines are BM25 Fusion (base retrieval), T5 Fusion as well as SciBERT Fusion.} 
    
    \label{tab:overall}
    \begin{tabular}{l|l|c c|c c}
    \hline
    \multirow{2}{*}{\textbf{Run ID}} & \multirow{2}{*}{\textbf{Method}} & \multicolumn{2}{c|}{\textbf{R1 (dev)}} & \multicolumn{2}{c}{\textbf{R2 (test)}} \\
    \cline{3-6} & & \textbf{NDCG@10} & \textbf{P@5} & \textbf{NDCG@10} & \textbf{P@5} \\
    \hline
     r2.fusion2 & \text{BM25 Fusion} & \textbf{0.6056} & \textbf{0.7200} & 0.5553 & 0.6800 \\
    covidex.t5 & \text{T5 Fusion} & 0.5124 & 0.6333 & 0.6250 & 0.7314 \\
    GUIR S2 run1 & \text{SciBERT Fusion} & 0.6032 & 0.6867 & \textbf{0.6251} & \textbf{0.7486}  \\
    \hline
     SparseDenseSciBert & \text{SciBERT + DAPT + DenseRetrieval$\text{}^*$} & \textbf{0.7424} & \textbf{0.8933} & \textbf{0.6772} & \textbf{0.7600}  \\
     ReInfoSelect & \text{SciBERT + DAPT + ContrastQG + ReInfoSelect$\text{}^*$} & 0.7134 & 0.8333 & 0.6259 & 0.6971 \\
      n.a. & \text{SciBERT + DAPT + ReInfoSelect} & 0.7061 & 0.8000 & 0.6210 & 0.6914 \\
     ContrastNLGSciBert & \text{SciBERT + DAPT + ContrastQG$\text{}^*$} & 0.6830 & 0.8467 & 0.6138 & 0.7314 \\
     n.a. & \text{SciBERT + DAPT} & 0.6775 & 0.7400 & 0.5880 & 0.6800  \\
     n.a. & \text{SciBERT} & 0.6598 & 0.7733 & 0.5828 & 0.6629 \\
    \hline
    \end{tabular}
\end{table*}

%% file: Sections/4_Experiment.tex
\section{Implementation Details.}
\label{sec:implement}
In this section, we describe the system's implementation details.

\textbf{Dataset.} The testing data of TREC-COVID Round 2 contains the May 1, 2020 version of the CORD-19 document set~\cite{wang2020cord} (59,851 COVID-related papers) and 35 queries written by biomedical professionals. Among these queries, the first 30 queries have been judged in the Round 1. In the experiment, we use TREC-COVID Round 1's annotated data as the development set (30 queries) and the medical MS MARCO~\cite{macavaney2020sledge} as the training data (78,895 queries).

\textbf{System Setup.} For data preprocessing, we concatenated title and abstract to represent each document and deleted stop words for all queries. Our system utilized the BM25 constructed by Anserini~\cite{yang2017anserini} as the base retrieval and adopted the dense retrieval implementation provided by Gao, et al.~\cite{gao2020complementing}. The neural ranker based on SciBERT~\cite{Beltagy2019SciBERT} was used in dense retrieval and reranking stages~\cite{macavaney2020sledge} with the learning rate of 2e-5 and the batch size of 32. We set the warm-up proportion as 0.1 and limited the maximum sequence length to 256. The NDCG@10 score on the development set is used to measure the convergence and is calculated every three training steps. Our system is based on PyTorch, and the training process it involves can be implemented on a GeForce RTX 2080 Ti.



\input{Diagrams/qid2ndcg}
\input{Diagrams/residual}

%% file: Diagrams/qid2ndcg.tex
\begin{figure*}[t]
    \centering
    \includegraphics[width=\textwidth]{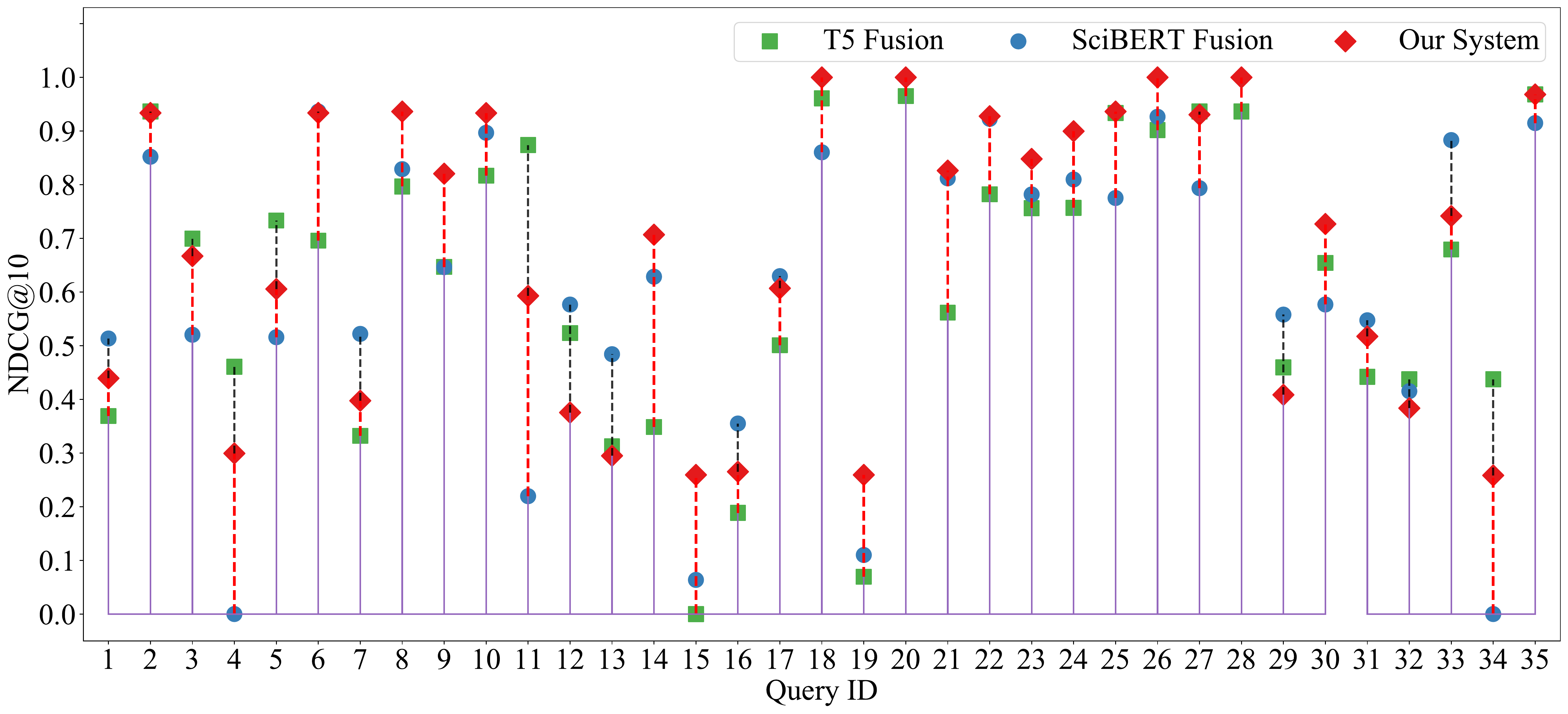}
    \caption{Round 2 testing result on each query of baselines and our system's best version (SciBERT + DAPT + Dense Retrieval). The X-axis denotes the Query ID in Round 2 of TREC-COVID, and Y-axis represents the NDCG@10 score. Noted that Queries 1-30 have been annotated in Round 1, and Queries 31-35 are newly added in Round 2.\label{fig:qid2ndcg}}
\end{figure*}



%% file: Diagrams/residual.tex
\begin{figure}[t]
\centering
\includegraphics[width=0.48\textwidth]{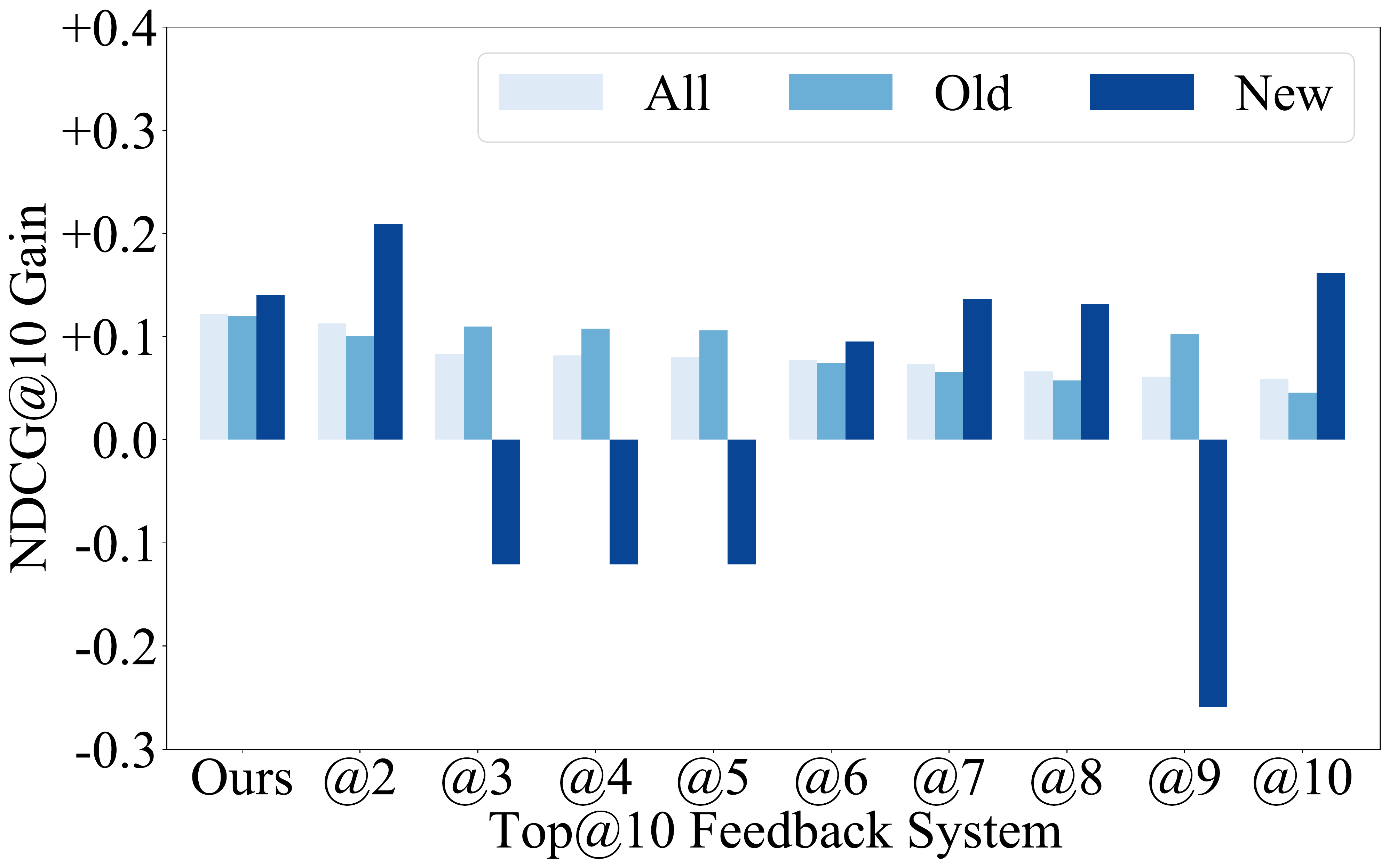}
\caption{The NDCG@10 gain of the top 10 feedback systems relative to BM25 Fusion system. `All' represents the average gain of all queries in TREC-COVID Round 2, `Old' and `New' mean the annotated queries in Round 1 and the newly added queries in Round 2, respectively.\label{fig:residual}}
\end{figure}

%% file: Sections/5_Evaluation.tex
\section{Evaluation Results}
\label{sec:evaluation}
This section presents evaluation results and hyperparameter studies.

\input{Diagrams/rankdepth}

\subsection{Overall Results}
Table~\ref{tab:overall} shows the overall performance of different models in the TREC-COVID task. Three top systems during Round 2 evaluation and several variants of our systems are compared.

Our system achieved the best performance in Round 2 of TREC-COVID. From our detailed experimental results, our method significant improves the ranking performance of SciBERT in the COVID domain. The domain-adaptive pretraining (DAPT) helps to improve SciBERT, which illustrates that learning the semantics of these new terminologies is crucial for language models. Then the system's performance has been further improved with about 6.5\% NDCG@10 gains by ContrastQG and ReInfoSelect. ContrastQG generates lots of pseudo relevance labels, which provides more training guidance for neural rankers in the specific domain. ReInfoSelect further boosts models with more fine-grained selected supervisions. The most significant improvement comes from the fusion of dense retrieval, where the P@5 score is increased by 11.8\%. This result shows that dense retrieval can significantly improve retrieval effectiveness by alleviating sparse retrieval's vocabulary mismatch problem.


\subsection{Hyperparameter Study}
Among all hyperparameters, we found the reranking depth significantly impacts the neural ranking model's effectiveness. As shown in Table~\ref{tab:rankdepth}, SciBERT's performance is significantly limited at the shallow reranking depth ($\leq$20), mainly caused by the low ranking accuracy of BM25. With the increase the reranking depth to 50 and 100, the neural ranker shows stable performance and achieves the best. Nevertheless, the reranking accuracy begins to drop as the depth continues to increase. The possible reason is that the neural ranker is not good enough to distinguish truly relevant documents when more noisy documents are included. 


\subsection{Query Analysis}
Figure~\ref {fig:qid2ndcg} shows the testing results of each query. The first 30 queries have been judged in Round 1, and others are newly added in Round 2 (query 31-35). Our system outperforms baselines on most queries with previous annotations. Besides, our system is also comparable to the T5 Fusion system on new queries and avoids the sharp drop of the SciBERT Fusion system (such as 34th query), which shows our system's robustness.


%% file: Diagrams/rankdepth.tex
\begin{table}[t]
    \centering
    \caption{Dev results of SciBERT with different reranking depth in Round 2 of TREC-COVID. The top 10 hole rate denotes the unlabeled proportion of the top 10 reranked results. \label{tab:rankdepth}}
    \begin{tabular}{c|c c|c}
     \hline
    \bf{Rerank Depth} & \bf{NDCG@10} & \bf{P@5} & \bf{Top 10 Hole Rate} \\
     \hline
     \text{20} & 0.6545 & 0.7429 & 0.03 \\
     \text{50} & \bf{0.6853} & \bf{0.7714} & 0.07 \\
     \text{100} & 0.6838 & 0.7543 & 0.12 \\
     \text{500} & 0.6044 & 0.6971 & 0.23 \\
     \text{1000} & 0.5826 & 0.6686 & 0.26 \\
    \hline
    \end{tabular}
\end{table}


%% file: Sections/6_NegativeRes.tex
\section{Failed Attempts}
\label{sec:attempts}
This section discusses some of our failed attempts and experience.

\textbf{Manual Labeling.} A straightforward approach to mitigate the label scarcity is to annotate more data within this domain manually. We recruited three medical students who compiled 50 COVID-related queries and assigned the relevance label to the top 20 documents retrieved by BM25 for each query. However, our annotations were not able to get good agreement with TREC-COVID's annotations.


\textbf{Corpus Filtering.} MacAvaney et al.~\cite{macavaney2020sledge} proposed to narrow the retrieval scale by filtering out the document published before 2020. Nevertheless, our analysis found that this method excluded more than 80\% of documents from the second round of corpus, dropping a large amount of useful COVID-related literature, such as SARS and MERS. Thus, we did not adopt this method in our system.

\textbf{Neural Reranker.} We also attempted two other neural ranking models besides SciBERT for document reranking, including BERT~\cite{devlin2019bert} and Conv-KNRM~\cite{dai2018convolutional}. Our experimental results show that BERT-Large has no obvious advantage over SciBERT-Base and Conv-KNRM performs the worst. The main reason for the poor performance of Conv-KNRM is that we did not use its subword version~\cite{hofstatter2019effect}, which led to a severe out-of-vocabulary problem.

\textbf{Fusion Attempts.} Two fusion methods have been tried to integrate dense retrieval into our system. One approach is to combine dense retrieval with BM25 in the base retrieval stage. The other is to fuse dense retrieval into SciBERT's reranking processing directly. The second method works better in our limited attempts.



\section{Concerns on Residual Evaluation}
\label{sec:concern}

This section discusses our observations about the residual collection evaluation used in the TREC-COVID task. In residual collection evaluation, test queries can be divided into \textit{old queries} and \textit{new queries}. The old queries have been annotated in previous rounds, but their annotated documents will be removed from the collection before scoring. TREC-COVID allows IR systems to use old queries' relevance judgments and classify such systems as feedback types.


Figure~\ref{fig:residual} shows the evaluation results of the top 10 feedback systems in Round 2 of TREC-COVID. Although these systems performed closely in overall scores, they showed significant differences in the old and new queries. E.g., the 2nd system's performance in the new query is greatly better than that in the old query. In contrast, some systems' ranking accuracy for the new query is considerably lower than in the old query, even worse than the base retrieval BM25 Fusion system, such as the 3rd-5th and 9th systems.

A powerful search system is desirable to achieve balanced performance on known and unknown queries. However, this result shows that the residual collection evaluation may bias towards seen queries, which are much easier in real production scenarios.
